\RequirePackage{fix-cm}
\documentclass[smallextended]{svjour3}       
\smartqed  
\usepackage{graphicx}
\usepackage{color}
\usepackage{bm,bbm,amssymb,amsmath}

\usepackage{multirow}
\newcommand{\hxx}{\hat{x}}
\newcommand{\ph}{\varphi}

\begin{document}

\title{Nuclei with up to $\boldsymbol{A=6}$ nucleons with artificial neural network wave functions} 


\author{Alex Gnech \and Corey Adams \and Nicholas Brawand \and Giuseppe Carleo \and Alessandro Lovato \and Noemi Rocco
}


\institute{ A. Gnech \at
                Theory Center, Jefferson Lab, Newport News, VA 23606, USA \\
                \email{agnech@jlab.org}
            \and
            C. Adams \at 
                Physics Division and Leadership Computing Facility, Argonne National Laboratory, Argonne, Illinois 60439, USA \\
                \email{corey.adams@anl.gov}
            \and
            N. Brawand \at
              Physics Division, Argonne National Laboratory, Argonne, Illinois 60439, USA \\
              \email{nbrawand@anl.gov}           
           \and
            G. Carleo \at 
                Institute of Physics, École Polytechnique Fédérale de Lausanne (EPFL), CH-1015 Lausanne, Switzerland \\
                \email{giuseppe.carleo@epfl.ch}
            \and
            A. Lovato \at 
                Physics Division and Computational Science Division, Argonne National Laboratory, Argonne, Illinois 60439, USA \\
                \email{lovato@anl.gov}
            \and
            N. Rocco \at
                Theoretical Physics Department, Fermi National Accelerator Laboratory, P.O. Box 500, Batavia, Illinois 60510, USA \\
                \email{nrocco@fnal.gov}
}

\date{Received: date / Accepted: date}

\maketitle

\begin{abstract}
The ground-breaking works of Weinberg have opened the way to calculations of atomic nuclei that are based on systematically improvable Hamiltonians. Solving the associated many-body Schr\"odinger equation involves non-trivial difficulties, due to the non-perturbative nature and strong spin-isospin dependence of nuclear interactions. Artificial neural networks have proven to be able to compactly represent the wave functions of nuclei with up to $A=4$ nucleons. In this work, we extend this approach to $^6$Li and $^6$He nuclei, using as input a leading-order pionless effective field theory Hamiltonian. We successfully benchmark their binding energies, point-nucleon densities, and radii with the highly-accurate hyperspherical harmonics method.

\end{abstract}

\section{Introduction}
\label{intro}
At the energy regime relevant for the description of atomic nuclei, the fundamental theory of strong interactions, quantum chromodynamics (QCD), becomes non-perturbative in its coupling constant. As a consequence, it is a formidable challenge to understand how nuclei directly emanate from the QCD Lagrangian. The ground-breaking works of Steven Weinberg~\cite{Weinberg:1990rz,Weinberg:1991um,Weinberg:1992yk} have opened the way to establishing nuclear effective field theories (EFTs) as the link between QCD and nuclear observables. Nuclear EFTs exploit the separation between the “hard” momentum scale ($M$, typically the nucleon mass) and the “soft” momentum scale ($Q$, typically the exchanged momentum). The active degrees of freedom at soft scale are hadrons whose interactions are constrained by the symmetries of QCD. Effective potentials and currents are then derived from the most general EFT-Lagrangian, and can be employed to make predictions for nuclear observables in a systematic expansion in $Q/M$~\cite{vanKolck2014}. To describe processes characterized by $Q \sim m_\pi$, the best suited approach is chiral-EFT which exploits the (approximate) chiral symmetry of QCD to derive consistent nuclear potentials and currents to estimate and reduce their uncertainties~\cite{Epelbaum:2008ga,Machleidt:2011zz}. For distances larger than $1/m_\pi$, $Q$ is smaller than both $m_\pi$ and $M$, so that pions can be integrated out and the nuclear interactions reduce to delta functions and their derivatives~\cite{Bedaque:2002mn}. This pionless-EFT is particularly useful when describing systems characterized by distances much larger than the two-nucleon scattering lengths–that is, the two-nucleon scattering amplitude at zero energy~\cite{vanKolck2014}. 

Nuclear EFTs, originally proposed by Steven Weinberg, are nowadays the main input to ``ab-initio'' many-body approaches, which are aimed at solving the many-body Schr\"odinger equation associated with the nuclear Hamiltonian with controlled approximations~\cite{Hergert:2020bxy}. Among them, continuum quantum Monte Carlo (QMC) methods, such as the variational Monte Carlo (VMC), Green’s function Monte Carlo (GFMC), and auxiliary-field diffusion Monte Carlo (AFDMC), are ideally suited to test the convergence of Weinberg's power counting and, more in general, the predictions of nuclear EFTs. These methods can accommodate ``bare'' potentials derived within nuclear EFTs, without further regularizing them and hence avoid the appearance of induced many-body forces. In addition, they have no problem in dealing with ``stiff'' forces, thereby enabling the exploration of a wide range of regulator values. Despite their success in describing the structure and dynamics of light nuclei~\cite{Carlson:2014vla}, QMC techniques face important challenges. The calculation of  spin-isospin dependent Jastrow correlations used in GFMC scales exponentially with the number of nucleons, limiting the applicability of these methods to nuclei with up to $A=12$ nucleons. On the other hand, the Hubbard Stratonovich transformations used by  AFDMC~\cite{Schmidt:1999lik} to treat larger nuclear systems~\cite{Piarulli:2019pfq,Lonardoni:2019ypg} are limited to somewhat simplified interactions~\cite{Gandolfi:2020pbj}. In addition, the use of wave functions that scale polynomially with the number of nucleons exacerbates the AFDMC fermion sign problem. Thus, extending continuum QMC calculations to medium-mass nuclei requires devising wave functions that exhibit a polynomial scaling with $A$ while still encompassing the vast majority of nuclear correlations.

Algorithms that take advantage of noisy intermediate-scale quantum devices are potentially groundbreaking alternatives, and their capabilities have already been demonstrated on prototypical nuclear problems~\cite{Dumitrescu:2018njn,Roggero:2018hrn,Roggero:2019srp}. An alternative class of approaches relies on the ability of artificial neural networks (ANNs) to compactly represent complex high-dimensional functions~\cite{carleo_machine_2019}.
The ANN variational representations of quantum mechanical wave functions have been introduced in Ref.~\cite{Carleo:2017} and have then since been applied to study ground-state properties and dynamics of several interacting lattice and continuum quantum systems~\cite{nomura_restricted_2017,Saito:2018b,Choo:2018,Nomura:2020,yoshioka:2019,nagy_variational_2019,vicentini:2019,hartmann_neural-network_2019,ferrari_neural_2019,Pfau:2019,Hermann:2019,Choo:2019}. In the domain of low-energy nuclear physics, Ref.~\cite{Keeble:2019bkv} has provided a proof-of-principle non-stochastic application of ANN to solve the Schr\"odinger equation of the deuteron with realistic Chiral-EFT interactions. Subsequently, the authors of Ref.~\cite{Adams:2020aax} have presented a VMC-ANN algorithm that extends the domain of applicability of ANN-based representations of the nuclear wave functions to nuclei with up to $A = 4$. An adaptive stochastic reconfiguration algorithm has been devised to efficiently train permutation-invariant ANNs and compute ground-state properties of $A\leq 4$ nuclei as they emerge from a leading-order pionless-EFT Hamiltonian.

In this work, we refine the VMC-ANN algorithm of Ref.~\cite{Adams:2020aax}, so that it now takes as input the permutation-invariant ANN pair-wise coordinates instead of single-particle ones. We also extend its applicability to open-shell nuclei with $A=6$ nucleons and validate its predictions against the highly-accurate hypershperical-harmonics (HH) few-body method~\cite{Kievsky:2008jpg}. We solve the Schr\"odinger equation for the pionless-EFT Hamiltonian of Ref.~\cite{Schiavilla:2021dun} and compute the binding energies and charge radii of $^2$H, $^3$H, $^3$He, $^4$He, $^6$He, $^6$Li and compare our results with experimental data. The paper is organized as follows. Section~\ref{sec:hamiltonian} is devoted to the pionless-EFT Hamiltonian that we use, the VMC-ANN and HH methods are described in Section~\ref{sec:methods}. Our results are presented and discussed in Section~\ref{sec:results}, while in Section~\ref{sec:conclusion} we draw our conclusions.

\section{Nuclear Hamiltonian}
\label{sec:hamiltonian}
We employ a nuclear Hamiltonian that is based on the tenet that the momentum scale relevant to model the structure of atomic nuclei is much smaller than the pion mass $m_\pi \simeq 140$ MeV. In this regime, pions are integrated out, hence the name pionless-EFT~\cite{Chen:1999tn,Bedaque:2002mn}, and the charge-independent (CI) nuclear interactions only consist of contact terms between two or more nucleons. At leading-order (LO) in the pionless-EFT expansion, the Hamiltonian is a sum of a non-relativistic kinetic-energy term plus nucleon-nucleon ($N\!N$) and three-nucleon ($3N$) contact potentials 
\begin{align}
H_{LO} &=-\sum_i \frac{{\vec{\nabla}_i^2}}{2m_N} 
+\sum_{i<j} v_{ij}+\sum_{i<j<k} V_{ijk}
\label{eq:ham}
\end{align}
In addition to the CI component, the two-nucleon interaction contains an electromagnetic (EM) contribution: $v= v^{\rm EM}+v^{\rm CI}_{\rm LO}$, where the full $v^{\rm EM}$ includes one- and two-photon Coulomb terms, the Darwin-Foldy term, vacuum polarization, and the magnetic moment interactions --- see Ref.~\cite{Wiringa:1994wb} for their full expressions. However, as in the AFDMC calculations of Ref.~\cite{Schiavilla:2021dun}, in the present work we only retain the Coulomb repulsion between finite-size (rather than point-like) protons. 

The LO pionless-EFT $N\!N$ interaction derived in Ref.~\cite{Schiavilla:2021dun} is required to only act in even partial waves. In coordinate space, it reads
\begin{equation}
v_{LO}^{CI}(r_{ij}) = C_{01} C_1(r_{ij})P_0^\sigma P_1^\tau + C_{10} C_0(r_{ij}) P_1^\sigma P_0^\tau 
\label{eq:vNN_LO}
\end{equation}
where $r_{ij}=|{\bf r}_i-{\bf r}_j|$ and $P_{0,1}^\sigma$ ($P_{0,1}^\tau$) are spin (isospin) projection operators on pair $ij$ with $S$ ($T$) equal to $0$ and $1$
\begin{equation}
P_0^\sigma = \frac{1 - \sigma_{ij}}{4},\quad
P_1^\sigma = \frac{3 + \sigma_{ij}}{4},\quad
P_0^\tau = \frac{1 - \tau_{ij}}{4},\quad
P_1^\tau = \frac{3 + \tau_{ij}}{4}\, .
\end{equation}
In the above equation we define $\sigma_{ij} = {\bm \sigma}_i\cdot{\bm \sigma}_j$ and $\tau_{ij} = {\bm \tau}_i\cdot{\bm \tau}_j$, with ${\bm \sigma}$ and ${\bm \tau}$ denoting the Pauli spin and isospin operator, respectively. The Gaussian cutoff functions
\begin{equation}
C_{\alpha}(r)=\frac{1}{\pi^{3/2}R_\alpha^3} e^{-(r / R_\alpha)^2}
\end{equation}
are introduced to regularize the contact interactions. In this work, we use model ``o'' of Ref.~\cite{Schiavilla:2021dun}, whose cutoff radii $R_0= 1.54592984$ fm and $R_1=1.83039397$ fm as well as the low-energy constants $C_{01}=-5.27518671$ fm$^2$ and $C_{10}= -7.04040080$ fm$^2$ have been adjusted by fitting the neutron-proton scattering lengths and effective radii in the singlet and triplet channel, and the deuteron binding energy. 

\begin{figure*}
\centering
  \includegraphics[width=0.75\textwidth]{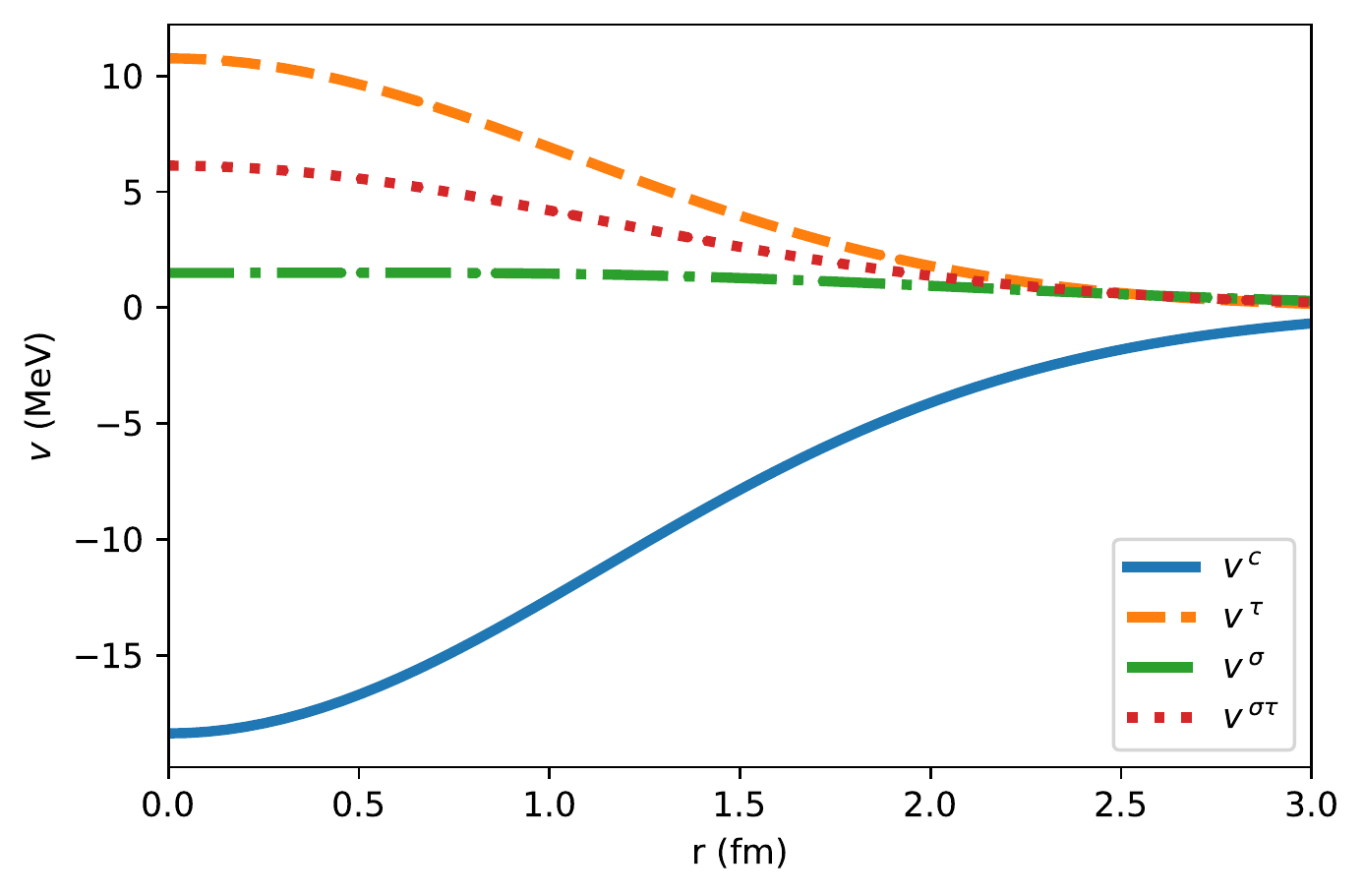}
\caption{Radial functions of the model ``o'' $N\!N$ potential at LO in the pionless-EFT expansion in the spin-isospin basis.}
\label{fig:v_NN}       
\end{figure*}

The potential of Eq.~\eqref{eq:vNN_LO} can be compactly written in the spin-isospin operator basis as  
\begin{align}
    v^{\rm CI}_{ij}= \sum_{p=1}^4 v^{p}(r_{ij})O^p_{ij}\, ,
    \label{eq:NN_op}
\end{align}
where $O^{p=1,4}_{ij}= ( 1, \tau_{ij}, \sigma_{ij}, \sigma_{ij}\tau_{ij} )$. The explicit expressions of the radial functions $v^p(r_{ij})$ are provided in Appendix~A of Ref.~\cite{Schiavilla:2021dun} and their radial dependence is displayed in Fig.~\ref{fig:v_NN}. 

In pionless-EFT, solving $A\geq 3$ nuclei with purely attractive LO two-nucleon potentials leads to their ``Thomas collapse''~\cite{Yang:2019hkn} when the regulator is taken to infinity. This pathological behavior is avoided promoting a contact $3N$ force to LO~\cite{Bedaque:1998kg}. In this work, we take a regularized $3N$ potential of the form
\begin{align}
    V_{ijk}= \frac{c_E}{f_\pi^4\Lambda_\chi}\frac{(\hbar c)^6}{\pi^3 R_3^6} \sum_{\rm cyc} e^{-(r_{ij}^2+r_{jk}^2)/R_3^2}
\end{align}
where $\Lambda_\chi=1$ GeV, $f_\pi= 92.4$ MeV is the pion decay constant and $\sum_{\rm cyc}$ stands for the cyclic permutation of indices $i, j$, and $k$. The LEC $c_E$ is fixed to reproduce the $^3$H binding energy, $B(^3H)= 8.475$ MeV, for a given value of the cutoff $R_3$. From the analysis of Ref.~\cite{Schiavilla:2021dun}, it emerges that the choice $R_3= 1.0$ fm  yields a satisfactory description of nuclear binding energies for several nuclei over a broad range of masses, up to $^{90}$Zr. Therefore, we adopt this value for the regulator and the corresponding $c_E= 1.0786$, as the Coulomb term between finite size protons does not contribute to the binding energy of $^3$H. 

\section{Few-body methods}
\label{sec:methods}

\paragraph{Variational Monte Carlo with ANN wave functions}
The variational Monte Carlo method seeks the ground-state of a given Hamiltonian by minimizing the energy expectation value
\begin{align}
\frac{\langle \Psi_V | H | \Psi_V \rangle}{\langle \Psi_V | \Psi_V \rangle} = E_V \geq E_0 
\label{eq:H_exp}
\end{align}
where $E_0$ is the exact ground-state energy: $H|\Psi_0\rangle = E_0 |\Psi_0\rangle$. The Metropolis-Hastings algorithm~\cite{Metropolis:1953am} is employed to sample the spatial and spin-isospin coordinates and carry out the multi-dimensional integration needed to evaluate the variational energy $E_V$.\\ 
In the same spirit as neural-network wave functions employed in condensed-matter, chemistry, and nuclear physics applications~\cite{nomura_restricted_2017,ferrari_neural_2019,Hermann:2019,Adams:2020aax}, we construct the ANN variational state as a product of mean-field state modulated by a flexible correlator factor 
\begin{align}
\langle R S |\Psi_V^{\textrm{ANN}} \rangle =  e^{\,\mathcal{U}(R,S)} \tanh[\mathcal{V}(R,S)] \langle R S |\Phi\rangle\,  .
\label{eq:psi_ANN}
\end{align}
In the above equation, $R\equiv\{\mathbf{r}_1,\dots,\mathbf{r}_A\}$ and $S\equiv\{\mathbf{s}_1,\dots,\mathbf{s}_A\}$ denote the set of single-particle spatial three-dimensional coordinates and the z-projection of the spin-isospin degrees of freedom $\mathbf{s}_i = \{s^z_i, t_i^z\}$, respectively.\\
The mean-field part of the wave function is expressed as a sum of Slater determinants of single-particle orbitals 
\begin{equation}
\langle R S |\Phi\rangle = \Big[\sum_n C_n \mathcal{A}_n[\phi_{\alpha_1}(\mathbf{r}_1,\mathbf{s}_1) \dots \phi_{\alpha_A}(\mathbf{r}_A,\mathbf{s}_A)]\Big]_{J^\pi T}\, ,
\end{equation}
$\mathcal{A}$ being the antisymmetrization operator and $C_n$ are the appropriate Clebsh-Gordan coefficients to reproduce the total angular momentum, total isospin, and parity $(J^\pi, T)$ for the specific nucleus of interest. The single-particle orbitals are given by
\begin{equation}
\phi_{\alpha_i}(\mathbf{r}_i,\mathbf{s}_i) = \mathcal{R}_{nl}(r_i) Y_{ll_z}(\hat{r}_i) \chi(s^z_i) \eta (t_i^z)\, ,
\end{equation}
where $\mathcal{R}_{nl}(r_i)$ are the radial functions parametrized by  feed-forward ANNs, $Y_{ll_z}(\hat{r}_i)$ are the spherical harmonics, and $\chi(s^z_i)$ and $\eta (t_i^z)$ are the spinors describing the spin and isospin of the single-particle state.  To automatically remove the spurious center of mass contribution from the kinetic energy, the spatial coordinates are replaced by $\mathbf{r}_i \to \mathbf{r}_i - \mathbf{R}_{\rm CM}$, with $\mathbf{R}_{\rm CM} = \frac{1}{A}\sum_i \mathbf{r}_i$ being the center of mass coordinate~\cite{Massella:2018xdj}.\\
Similarly to Ref.~\cite{Adams:2020aax}, the correlations factors $\mathcal{U}(R,S)$ and $\mathcal{V}(R,S)$ are expressed in terms of permutation-invariant ANNs based on the Deep Sets architecture~\cite{Zaheer:2017,Wagstaff:2019}. Instead of the single-particle inputs used in Ref.~\cite{Adams:2020aax}, in this work we map the coordinates of each pair separately to a latent-space representation --- pair-wise inputs were also used in the ANNs employed in Refs.~\cite{Pfau:2019,Hermann:2019}. A sum operation is then applied to destroy ordering information and ensure permutation invariance
\begin{equation}
\mathcal{F}(R,S) = \rho_\mathcal{F}\left(\sum_{i \neq j} \phi_\mathcal{F}(\mathbf{r}_i, \mathbf{s}_i,\mathbf{r}_j, \mathbf{s}_j)\right)\, \quad \mathcal{F} = \mathcal{U}, \mathcal{V}\, .
\end{equation}
When single-particle coordinates are used as input, correlations are generated in the latent space by the net $\rho_\mathcal{F}$. On the other hand, using the pair coordinates builds correlations already in the real space. Note that the conventional two-body Jastrow ansatz can be recovered if the latent space is taken to be one-dimensional and $\rho_\mathcal{F}$ is the identity function $\rho_F(x)=x$.\\
Both $\phi_\mathcal{U}$ and $\rho_\mathcal{U}$ are represented by real-valued ANNs comprised of four fully connected layers with 32 nodes each, while $\phi_\mathcal{V}$ and $\rho_\mathcal{V}$ are made of two fully connected layers. The size of the latent spaces for $\mathcal{U}$ and $\mathcal{V}$ is also taken to be 32-dimensional. On the other hand, the ANN representing $\mathcal{R}_{nl}$ is made of two fully connected layers, with 32 nodes each. The calculation of the kinetic energy requires using differentiable activation functions. We find that $\tanh$ and softplus~\cite{Dugas:2000} yield fully consistent results.\\
Since the parameters of the network are randomly initialized, in the initial phases of the training, during the Metropolis walk, the nucleons can drift away from $\mathbf{R}_{\rm CM}$. To control this behavior, a Gaussian function is added to the single-particle radial functions to confine the nucleons within a finite volume $\mathcal{R}(r_i) \to \mathcal{R}(r_i) e^{-\alpha r_i^2}$, and we take $\alpha = 0.04$. \\
The stochastic reconfiguration algorithm~\cite{Sorella:2005} is employed to minimize the energy expectation value of Eq.~\eqref{eq:H_exp} and find the optimal set of variational parameters. To speed-up the convergence of the training procedure, we adopt the adaptive learning rate algorithm discussed in the supplemental material of Ref.~\cite{Adams:2020aax}. 

\paragraph{Hypershpherical harmonics}
The hyperspherical harmonic (HH) method has been used to study nuclei
of $A\leq4$~\cite{Kievsky:2008jpg,Marcucci:2019fphy} and has been recently developed to treat also nuclei with up to $A=6$ nucleons~\cite{Gnech:2020prc}. In the HH the center of mass motion of the nucleons is decoupled and the remaining $N=A-1$ internal spatial configurations are given through the Jacobi coordinates,
defined as
\begin{equation}
      \label{eq:jac1}
  \boldsymbol{x}_{N-j+1} = \sqrt{\frac{2 j}{j+1} }
  \Bigl [\boldsymbol{r}_{j+1} - \frac{1}{j}\sum_{i=1}^j
    \boldsymbol{r}_i \Bigr ]\ ,
  \qquad j=1,\ldots,N\ ,
\end{equation}
where $\boldsymbol{r}_i$ is the position of the $i$-th particle. The hyperspherical coordinate are given by
  \begin{equation}
    \rho=\sqrt{\sum_{i=1}^N x_{i}^2}\,\qquad,\Omega_{N}=\{\hxx_{1},\cdots,\hxx_{N},
    \ph_{2},\cdots,\ph_{N}\}\,,\label{eq:romega}
  \end{equation}
  where $\cos \ph_{i}=\frac{x_{i}}{\sqrt{x_{1}^2+\dots+x_{i}^2}}$ for
  $i=2,\dots,N$.
  By rewriting the kinetic energy operator using the hyperspherical coordinates
  an operator $\Lambda_N^2(\Omega_N)$ appears~\cite{Kievsky:2008jpg}. The eigenfunction of this operator are the so-called hyperspherical harmonics which we indicate with
  ${\cal Y}^{KLM_L}_{[K]}(\Omega_N)$ where $K$ is the eigenvalue of the operator
  $\Lambda_N^2(\Omega_N)$, $L$ the total angular momentum, $M_L$ the projection of the total angular momentum along $z$ and with $[K]$ we indicate all the remaining quantum numbers which uniquely identify the HH state. The nuclear wave function contains also spin and isospin degrees of freedom. The spin(isospin) is added through the function
  $\chi^{SM_S}_{[S]}$($\chi^{TM_T}_{[T]}$), where $S$($T$) is the total spin(isospin)
  and $M_S$($M_T$) the projection of the spin(isospin) on the $z$ direction.
  The formal expression for the HH+spin+isospin function
  can be found in Ref.~\cite{Kievsky:2008jpg} for $A=3,4$ and~\cite{Gnech:2020prc} for $A=6$. The basis is then written as
   \begin{equation}
    Y^{KLSJJ_zTM_T}_{[\alpha]}=\left[{\cal Y}^{KLM_L}_{[K]}(\Omega_N)\chi^{SM_S}_{[S]}\right]_{JJ_z}\chi^{TM_T}_{[T]}\,.
    \label{eq:hhst-basis}
  \end{equation}
  The wave function of an $A$-body bound state, having total angular momentum $J$, $J_z$ and parity $\pi$, and third component of the total isospin $M_T$, can be decomposed as a sum of Faddeev-like amplitudes as:
  \begin{equation}
    \Psi_A=\sum_{p=1}^{N_p}\psi(\boldsymbol{x}_1^{(p)},\dots,\boldsymbol{x}_N^{(p)})\,,
  \end{equation}
  where the sum $p$ runs over the $N_p$ even permutations of the particles.  Note that by exploiting the sum over the permutation it is possible to select term of the HH basis expansion which are anti-symmetric using only the quantum numbers. By using the HH+spin+isospin anti-symmetrized basis in Eq.~(\ref{eq:hhst-basis}) the wave function can be rewritten as
  \begin{equation}
    \Psi_A=\sum_{l,\alpha}c_{l,\alpha}f_l(\rho)\sum_{p=1}^{N_p}Y^{KLSJJ_zTM_T}_{[\alpha]}(\Omega_N^{(p)})\,.
  \end{equation}
 Here the function $f_l(\rho)$ is defined in terms of Laguerre polynomials multiplied by an exponential and with $\alpha$ we indicate the sum over all the quantum numbers which identify an anti-symmetric HH+spin+isospin state.
  The coefficients $c_{l,\alpha}$ are unknown. To determine them we solve the eigenvalue-eigenvector problem derived from the Rayleigh-Ritz variational principle.
The obtained binding energies are expected to be accurate at the level of 1 keV
and 10 keV for the three- and four-nucleon systems, respectively. The HH basis cannot reach a comparable accuracy for $A=6$ nuclei, and the extrapolation technique described is in Refs.~\cite{Gnech:2020prc,Gnech:2021prc} is needed to  obtain the binding energies and charge radii of $^6$Li and $^6$He.

\section{Results and Discussion}
\label{sec:results}
The binding energies and charge radii of $^2$H, $^3$H, $^3$He, $^4$He, $^6$He, and $^6$Li obtained with the VMC-ANN and HH methods are listed in Table~\ref{tab:res_NN}. For nuclei with $A\geq 3$, we separately show the results obtained with the two-body force alone ($N\!N$) and with the full LO Hamiltonian of Eq.~\eqref{eq:ham} that includes the three-nucleon interaction ($3N$). The expectation value of the charge radius is derived from the point-proton radius using the relation:
\begin{align}
	\left\langle r_{\rm ch}^2\right\rangle=
	\left\langle r_{\rm pt}^2\right\rangle+
	\left\langle R_p^2\right\rangle+
	\frac{A-Z}{Z}\left\langle R_n^2\right\rangle+
	\frac{3}{4m_p^2},
	\label{eq:rch}
\end{align}
where $r_{\rm pt}$ is the calculated point-proton radius, $\left\langle R_p^2\right\rangle=0.770(9)\,\rm{fm}^2$~\cite{ParticleDataGroup:2012pjm} the proton radius, 
$\left\langle R_n^2\right\rangle=-0.116(2)\,\rm{fm}^2$~\cite{ParticleDataGroup:2012pjm} the neutron radius, and $(3)/(4m_p^2)\approx0.033\,\rm{fm}^2$ the Darwin-Foldy correction~\cite{Friar:1997js}. The point-proton radius is calculated as
\begin{align}
	\left\langle r_{\rm pt}^2\right\rangle=\frac{1}{Z}\big\langle\Psi\big|\sum_i P_{p} |\mathbf{r}_i-\mathbf{R}_{\rm cm}|^2\big|\Psi\big\rangle,
\end{align}
where $Z$ is the number of protons and $P_{p}=(1 + \tau_{z_i})/2$ is the proton projector operator.

\begin{table*}[htb]
\setlength{\tabcolsep}{3.9pt}
\centering
\caption{Ground-state energies and charge radii for selected $A\leq 6$ nuclei obtained from the ANN and HH methods using as input the leading-order pionless-EFT Hamiltonian with and without the $3N$ force. We report also the experimental binding-energies from
Ref.~\cite{NNDC2021} and the charge radius taken from Refs.~\cite{CODATA2021,Amroun:1994npa,Morton:2006pra,Krauth:2021nat,Wang:2004prl,Puchalski:2013prl}.}
\label{tab:res_NN}  
\begin{tabular}{c c c c c c c c}
\hline\hline
\multirow{2}{*}{Nucleus} & \multirow{2}{*}{Potential} & \multicolumn{2}{c}{ANN}               & \multicolumn{2}{c}{HH} & \multicolumn{2}{c}{Exp.}                \\
\noalign{\smallskip}
        &           & $E\,(\rm MeV)$ & $r_{\rm ch}\,(\rm fm)$ & $E\,(\rm MeV)$ & $r_{\rm ch}\,(\rm fm)$ & $E\,(\rm MeV)$ & $r_{\rm ch}\,(\rm fm)$ \\     
\noalign{\smallskip}\hline\noalign{\smallskip}
$^2$H  & $N\!N$ & $-2.242(1)$   & $2.120(5)$ & $-2.242$  & $2.110(2)$ & $-2.225$ & $2.128$\\
\noalign{\smallskip}\hline\noalign{\smallskip}                                                               
\multirow{2}{*}{$^3$H} & $N\!N$& $-9.511(1)$   & $1.658(4)$ & $-9.744$  & $1.656(4)$ & \multirow{2}{*}{$-8.475$} & \multirow{2}{*}{$1.755(86)$}  \\
        & $3N$  & $-8.232(1)$   & $1.750(3)$ & $-8.475$  & $1.747(6)$ & &  \\
\noalign{\smallskip}\hline\noalign{\smallskip}                                                  
\multirow{2}{*}{$^3$He} & $N\!N$& $-8.800(1)$   & $1.845(3)$ & $-9.035$  & $1.848(6)$ & 
\multirow{2}{*}{$-7.718$} & \multirow{2}{*}{$1.964(1)$}  \\
        & $3N$  & $-7.564(1)$   & $1.961(3)$ & $-7.811$  & $1.969(8)$ & & \\
\noalign{\smallskip}\hline\noalign{\smallskip}
\multirow{2}{*}{$^4$He} & $N\!N$& $-36.841(1)$   & $1.484(3)$ & $-37.06$  & $1.485(4)$ &
\multirow{2}{*}{$-28.30$} & \multirow{2}{*}{$1.678$}  \\
        & $3N$  & $-27.903(1)$   & $1.643(2)$ & $-28.17$  & $1.646(4)$ & & \\
\noalign{\smallskip}\hline\noalign{\smallskip}
\multirow{2}{*}{$^6$He} & $N\!N$& $-37.25(4)$  & $1.895(2)$ & $-37.96(8)$  & $1.71(1)$ &
\multirow{2}{*}{$-29.27$} & \multirow{2}{*}{$2.05(1)$}  \\
        & $3N$  & $-27.46(2)$   & $>4.89(1)$ & $-27.41(8)$  & $>2.73$ & &\\
\noalign{\smallskip}\hline\noalign{\smallskip}
\multirow{2}{*}{$^6$Li} & $N\!N$& $-42.04(1)$   & $2.248(3)$ & $-42.51(5)$  & $2.09(2)$ &
\multirow{2}{*}{$-31.99$} & \multirow{2}{*}{$2.54(3)$}  \\
        & $3N$  & $-30.82(3)$   & $3.049(2)$ & $-31.00(8)$  & $>2.74$ & & \\
\noalign{\smallskip}\hline                                                        
\hline
\end{tabular}
\label{tab:afdmc-gfmc}
\end{table*}

We begin our comparison with the $^2$H nucleus. Similarly with the findings of Refs.~\cite{Keeble:2019bkv,Adams:2020aax}, the VMC-ANN method converges to the same binding energy as the HH. Because of the missing EM contributions in the $N\!N$ potential, both values are slightly more bound than the experimental datum. The VMC-ANN and HH charge radii are compatible with each other within errors and only marginally smaller than the experimental value.  

VMC-ANN underbinds $A=3$ nuclei by around $0.25$ MeV with respect to the HH method, for both the $N\!N$ and the $N\!N + 3N$ Hamiltonians. As noted in Ref.~\cite{Adams:2020aax}, these small discrepancies reflect the main limitation of the wave function ansatz of Eq.~\eqref{eq:psi_ANN}, i.e. the impossibility for the ANN correlator to compensate for the zeros of the mean-field part of the wave function $\langle R S | \Phi \rangle$. Nevertheless, the VMC-ANN and HH charge radii are very close to each other and, once the $3N$ force is included in the Hamiltonian, they are compatible with experiments. We also note that the HH binding energy of $^3$H agrees by construction with the experimental value, while the $^3$He differs from it by $\sim 0.1$ MeV. This small discrepancy might well be covered accounting for the neutron-proton mass difference and by using the full $v^{\rm EM}$ interaction rather than the simple Coulomb repulsion between finite-size protons.

\begin{figure*}
\centering
  \includegraphics[width=0.75\textwidth]{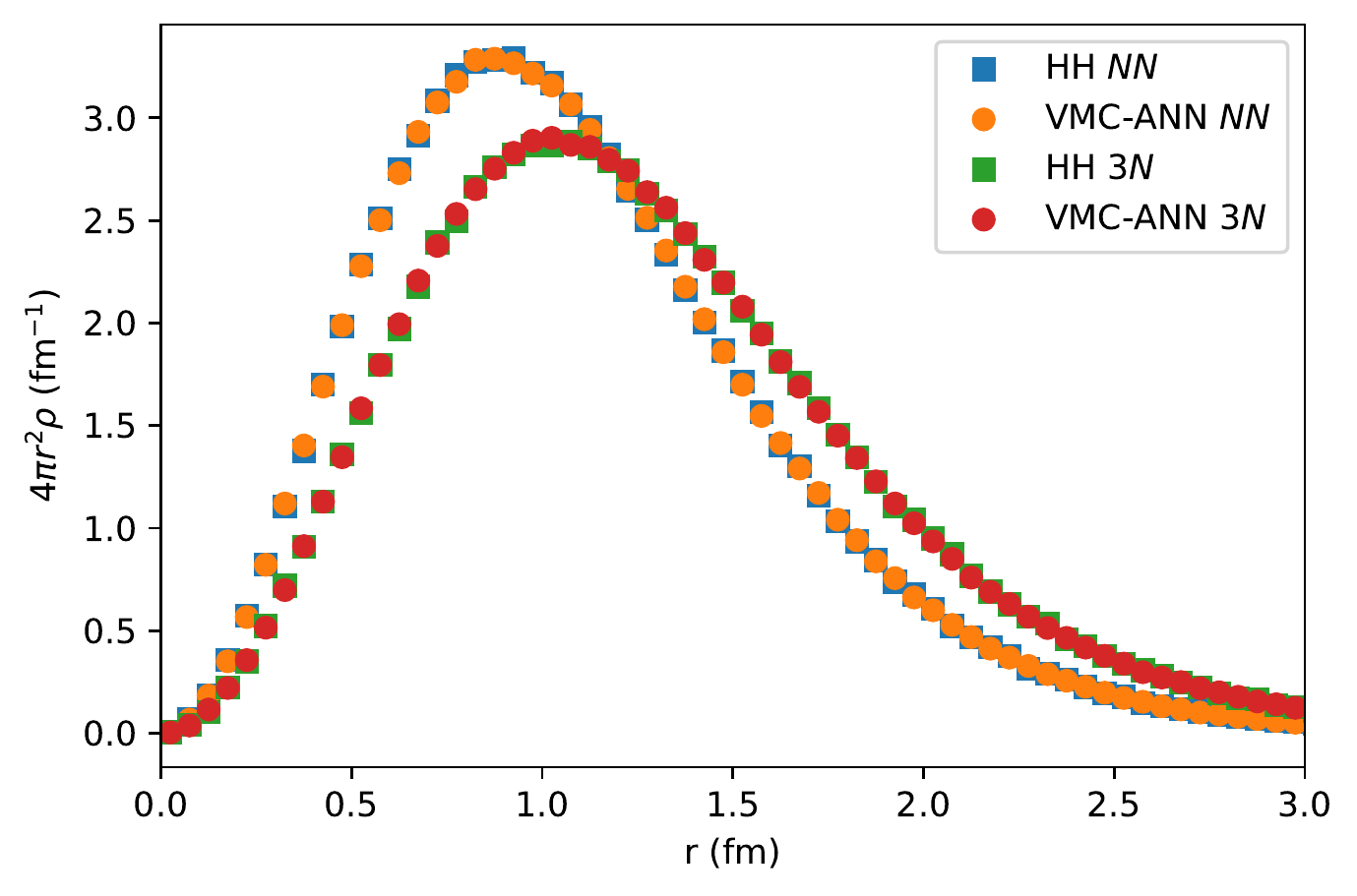}
\caption{VMC-ANN and HH point-nucleon density of $^4$He as obtained using as input the $NN$ force only or the full $NN$ + $3N$ Hamiltonian.}
\label{fig:rho_4he}     
\end{figure*}

Similarly to the $A=3$ case, the VMC-ANN method yields a ground-state energy of $^4$He that is $\sim 0.2$ above the HH value, whether or not the $3N$ force is included in the Hamiltonian. However, the repulsive features of the $3N$ potential are essential to bring the predicted binding energies and charge radii closer to their experimental values. To better illustrate this point, in Fig.~\ref{fig:rho_4he} we show the VMC-ANN and HH point-nucleon density of $^4$He as obtained with and without the $3N$ force. There is an excellent agreement between the two methods, corroborating once again the accuracy of ANNs in representing quantum-mechanical wave function of light nuclei. As expected, the $3N$ potential pushes nucleons further away from their CM, broadening the single-particle density and enlarging the charge radius of the nucleus. 

\begin{figure*}
\centering
  \includegraphics[width=0.75\textwidth]{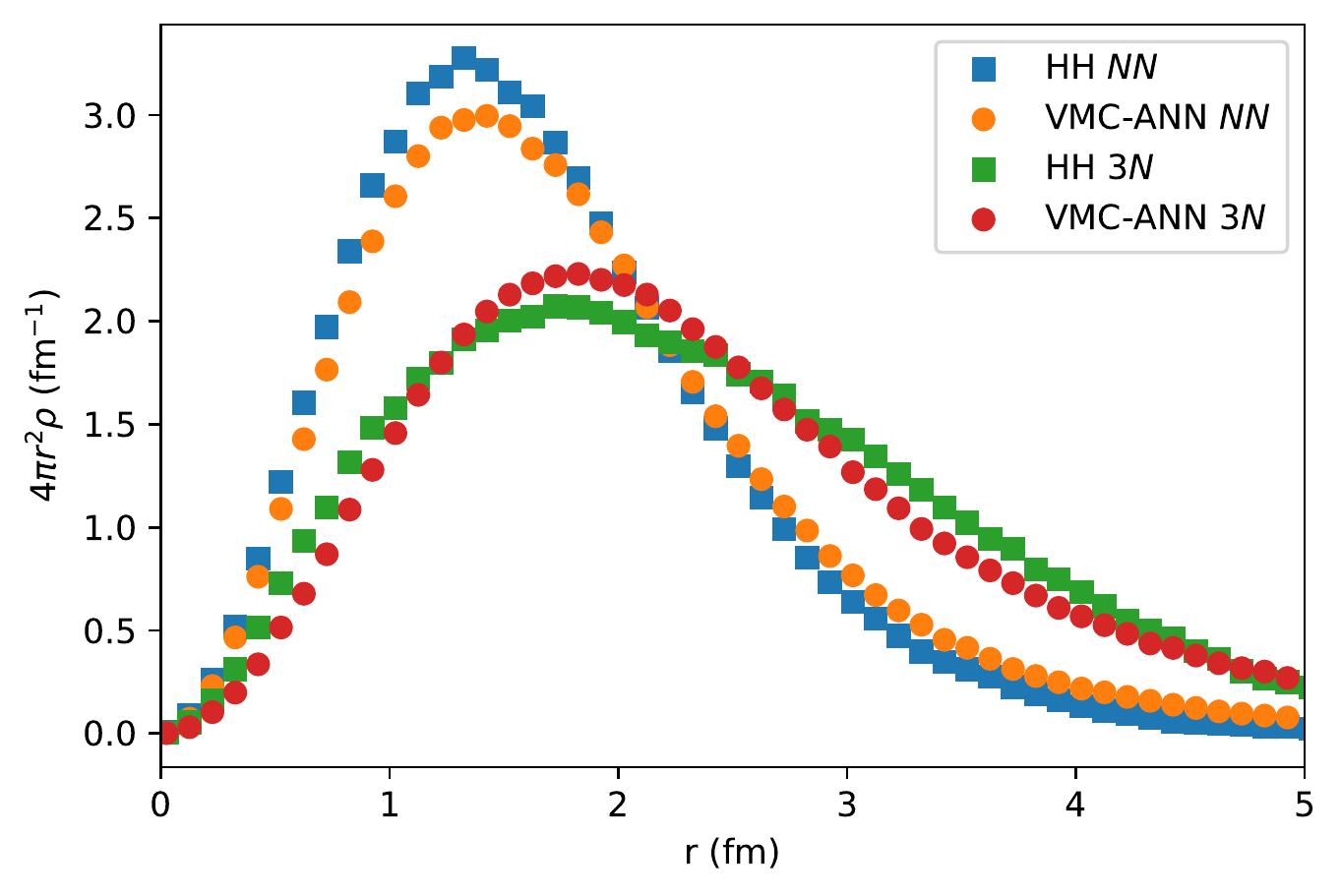}
\caption{VMC-ANN and HH point-nucleon density of $^6$Li as obtained using as input the $NN$ force only or the full $NN$ + $3N$ Hamiltonian.}
\label{fig:rho_6li}     
\end{figure*}

When the $N\!N$ only is taken as input, VMC-ANN produces $^6$He and $^6$Li wave functions that are stable against breakup into $^4$He plus two neutrons and $^4$He plus deuteron, respectively. This is a highly non-trivial results, as sophisticated VMC calculations of light nuclei that use conventional two- and three-body Jastrow correlations generally fail to get a binding energy for $^6$Li that is below the sum of the one of $^4$He and the deuteron~\cite{Wiringa:2021p}. It remains to be understood whether VMC-ANN would be able to get a stable $^6$Li once more realistic nuclear interactions that include a tensor component are employed. In the $N\!N$ case, the VMC-ANN energies are $\sim 0.5$ MeV less bound than those obtained using the HH approach, corresponding to less than $0.1$ MeV per nucleon. The latter value is not dissimilar from the one we get comparing the VMC-ANN and HH binding energies of $^3$H and $^3$He. \\
The VMC-ANN charge radii are appreciably larger than the HH ones for both $^6$He and $^6$Li nuclei, which are nevertheless smaller than the experimental values. This behavior is reflected in the point-nucleon density of $^6$Li, displayed in Fig.~\ref{fig:rho_6li}. The VMC-ANN distribution extends to larger distances than the HH one. This is partly due the poor quality of the HH in reproducing the long range behaviour of the $A=6$ wave functions, which translates in a very slow convergence rate for the charge radii as function of $K$ --- see Ref.~\cite{Gnech:2020prc} for a complete discussion. The relatively small value of $K$ employed in computing $A=6$ nuclei does not allow us to safely extrapolate the computed charge radii, which tend to be underestimated. \\
Analogous to  Fig.~\ref{fig:rho_4he}, the $3N$ interactions broaden the point-nucleon density distributions, depleting their value near $r=0$ and dramatically increasing the charge radius well above its experimental value. In fact, both VMC-ANN and HH calculations with the $3N$ potential yield a $^6$Li that is only barely bound against $^4$He plus deuteron breakup. On the other hand, neither VMC-ANN nor HH calculations that include the $3N$ potential yield a stable $^6$He, as its binding energy is above the one of $^4$He. Hence, its charge radius becomes larger and larger as the wave function keeps extending to larger distances from $\mathbf{R}_{\rm CM}$. This behavior is reminiscent to the one described in Ref.~\cite{Contessi:2017rww} for the $^{16}$O nucleus, which was unbound with respect to breakup into four alpha clusters.  

\section{Conclusion}
\label{sec:conclusion}
The development of QMC methods is instrumental for testing predictions and guiding the development of nuclear EFTs, originally proposed by Steven Weinberg. Currently available approaches are either limited to light nuclear systems or to simplified interactions. Novel VMC algorithms based on ANN representation of quantum-mechanical wave functions can potentially overcome these limitations and allow one to make predictions across the nuclear chart using a variety of EFT interactions and currents. 

In this work, we extended the reach of the VMC-ANN algorithm introduced in Ref.~\cite{Adams:2020aax} to compute the binding energies and charge radii of $A\leq 6$ nuclei as they emerge from the LO pionless-EFT Hamiltonian of Ref.~\cite{Schiavilla:2021dun}. The VMC-ANN architecture was made more efficient by using as input the pair-wise coordinates of the nucleons instead of the single-particle ones. In addition, treating open-shell $A=6$ nuclei has required summing multiple Slater determinant in the mean-field part of the wave function, which is also expressed in terms of ANNs.

We compare the VMC-ANN binding energies and charge radii of $A\leq 6$ nuclei with the highly accurate HH method. Performing benchmark calculations among different numerical methods corroborates our confidence in the accuracy of the nuclear Schr\"odinger equation solution~\cite{Kamada:2001tv,Maris:2013rgq,Piarulli:2019pfq}. This step is critical to carry out meaningful comparisons of nuclear EFTs' predictions against experimental data, as it helps to disentangle the uncertainties associated with solving the many-body problem from those pertaining the EFT-based modeling of nuclear dynamics.  

There is perfect agreement between the two methods for the simplest $^2$H nucleus, in terms of both binding energies and charge radii. On the other hand, VMC-ANN underbinds ${}^3$H, ${}^3$He, and ${}^4$He nuclei by $\sim0.2$ MeV with respect to HH. We ascribe the reason for this behavior to deficiencies in the mean-field part of the wave function that cannot be remedied by the ANN correlator. Nevertheless, the charge radii, and point-nucleon densities distributions for these $A\leq 4$ nuclei obtained within the VMC-ANN and HH methods are in excellent agreement, hence demonstrating the flexibility of the ANNs in representing the wave function of light nuclei. 

The difference in binding energy per nucleon between the VMC-ANN and HH methods is about $\sim0.1$ MeV for both $^6$He and $^6$Li; not dissimilar to the one observed for $A\leq4$ nuclei. The comparison of the VMC-ANN and HH charge radii and the point-nucleon for $A=6$ nuclei presents somewhat bigger discrepancies than in the $A\leq4$ case. The reason for this behavior is likely twofold. On the one hand, the HH expansion exhibits a slow convergence rate when it comes to reproducing the long-range behavior of the wave function. On the other one, VMC-ANN slightly underbinds these nuclei.

From our analysis, it appears that the $^6$He nucleus is not stable against breakup into $^4$He plus two neutrons at LO in the pionless-EFT expansion, at least for the values of the regulator and LECs that we used. The chief advantage of the Weinberg's nuclear EFT framework is that it contains the diagnostic tools to assess its convergence. To determine whether $^6$He is bound within pionless-EFT, in addition to using different values of the regulator and LECs, we will employ the next-to-leading (NLO) Hamiltonian of Ref.~\cite{Schiavilla:2021dun} and carry out a rigorous uncertainty quantification on the line of Refs.~\cite{Furnstahl:2014xsa,Melendez:2019izc}. Concurrently, we will further develop the accuracy of the VMC-ANN method and, thanks to its favorable polonomyal scaling with $A$, we will apply it to larger nuclear systems. The latter point is critical for studying the convergence and the predictive power of both chiral-EFT and pionless-EFT across the nuclear chart. 

\begin{acknowledgements}
Useful discussions with R. Schiavilla and R. B. Wiringa are gratefully acknowledged. The present research is supported by the U.S. Department of Energy, Office of Science, Office of Nuclear Physics, under contracts DE-AC05-06OR23177 (A.G.),  DE-AC02-06CH11357, by the NUCLEI SciDAC program (A.L., N.B.) and by Fermi Research Alliance, LLC under Contract No. DE-AC02-07CH11359 with the U.S. Department of Energy, Office of Science, Office of High Energy Physics (N.R.). A.L. and N.B. were also supported by DOE Early Career Research Program and Argonne LDRD awards. A.L acknowledges funding from the INFN grant INNN3, and from the European Union's Horizon 2020 research and innovation programme under grant agreement No 824093.
This research used resources of the Argonne Leadership Computing Facility, which is a DOE Office of Science User Facility supported under Contract DE-AC02-06CH11357.  The calculations were performed using resources of the Laboratory Computing Resource Center of Argonne National Laboratory, the National Energy Research Supercomputer Center (NERSC), and through a CINECA-INFN agreement that provides access to resources on MARCONI at CINECA. 
\end{acknowledgements}

\bibliographystyle{spphys}       
\bibliography{biblio.bib}   

\begin{thebibliography}{10}
\providecommand{\url}[1]{{#1}}
\providecommand{\urlprefix}{URL }
\expandafter\ifx\csname urlstyle\endcsname\relax
  \providecommand{\doi}[1]{DOI \discretionary{}{}{}#1}\else
  \providecommand{\doi}{DOI \discretionary{}{}{}\begingroup
  \urlstyle{rm}\Url}\fi

\bibitem{Weinberg:1990rz}
S.~Weinberg, Phys. Lett. \textbf{B251}, 288 (1990).
\newblock \doi{10.1016/0370-2693(90)90938-3}

\bibitem{Weinberg:1991um}
S.~Weinberg, Nucl. Phys. \textbf{B363}, 3 (1991).
\newblock \doi{10.1016/0550-3213(91)90231-L}

\bibitem{Weinberg:1992yk}
S.~Weinberg, Phys. Lett. \textbf{B295}, 114 (1992).
\newblock \doi{10.1016/0370-2693(92)90099-P}

\bibitem{vanKolck2014}
U.~{van Kolck}, \emph{{Effective Field Theories of Loosely Bound Nuclei}}
  (2014), vol. 879, p. 123.
\newblock \doi{10.1007/978-3-642-45141-6\_4}

\bibitem{Epelbaum:2008ga}
E.~Epelbaum, H.W. Hammer, U.G. Meissner, Rev. Mod. Phys. \textbf{81}, 1773
  (2009).
\newblock \doi{10.1103/RevModPhys.81.1773}

\bibitem{Machleidt:2011zz}
R.~Machleidt, D.~Entem, Phys. Rept. \textbf{503}, 1 (2011).
\newblock \doi{10.1016/j.physrep.2011.02.001}

\bibitem{Bedaque:2002mn}
P.F. Bedaque, U.~van Kolck, Ann. Rev. Nucl. Part. Sci. \textbf{52}, 339 (2002).
\newblock \doi{10.1146/annurev.nucl.52.050102.090637}

\bibitem{Hergert:2020bxy}
H.~Hergert, Front. in Phys. \textbf{8}, 379 (2020).
\newblock \doi{10.3389/fphy.2020.00379}

\bibitem{Carlson:2014vla}
J.~Carlson, S.~Gandolfi, F.~Pederiva, S.C. Pieper, R.~Schiavilla, K.~Schmidt,
  R.~Wiringa, Rev. Mod. Phys. \textbf{87}, 1067 (2015).
\newblock \doi{10.1103/RevModPhys.87.1067}

\bibitem{Schmidt:1999lik}
K.~Schmidt, S.~Fantoni, Phys. Lett. B \textbf{446}, 99 (1999).
\newblock \doi{10.1016/S0370-2693(98)01522-6}

\bibitem{Piarulli:2019pfq}
M.~Piarulli, I.~Bombaci, D.~Logoteta, A.~Lovato, R.~Wiringa, Phys. Rev. C
  \textbf{101}(4), 045801 (2020).
\newblock \doi{10.1103/PhysRevC.101.045801}

\bibitem{Lonardoni:2019ypg}
D.~Lonardoni, I.~Tews, S.~Gandolfi, J.~Carlson, Phys. Rev. Res. \textbf{2},
  022033 (2020).
\newblock \doi{10.1103/PhysRevResearch.2.022033}

\bibitem{Gandolfi:2020pbj}
S.~Gandolfi, D.~Lonardoni, A.~Lovato, M.~Piarulli, Front. Phys. \textbf{8}, 117
  (2020).
\newblock \doi{10.3389/fphy.2020.00117}

\bibitem{Dumitrescu:2018njn}
E.~Dumitrescu, A.~McCaskey, G.~Hagen, G.~Jansen, T.~Morris, T.~Papenbrock,
  R.~Pooser, D.~Dean, P.~Lougovski, Phys. Rev. Lett. \textbf{120}(21), 210501
  (2018).
\newblock \doi{10.1103/PhysRevLett.120.210501}

\bibitem{Roggero:2018hrn}
A.~Roggero, J.~Carlson, Phys. Rev. C \textbf{100}(3), 034610 (2019).
\newblock \doi{10.1103/PhysRevC.100.034610}

\bibitem{Roggero:2019srp}
A.~Roggero, A.~Baroni, Phys. Rev. A \textbf{101}(2), 022328 (2020).
\newblock \doi{10.1103/PhysRevA.101.022328}

\bibitem{carleo_machine_2019}
G.~Carleo, I.~Cirac, K.~Cranmer, L.~Daudet, M.~Schuld, N.~Tishby,
  L.~Vogt-Maranto, L.~Zdeborová, Reviews of Modern Physics \textbf{91}(4),
  045002 (2019).
\newblock \doi{10.1103/RevModPhys.91.045002}.
\newblock \urlprefix\url{https://link.aps.org/doi/10.1103/RevModPhys.91.045002}

\bibitem{Carleo:2017}
G.~Carleo, M.~Troyer, Science \textbf{355}(6325), 602 (2017).
\newblock \doi{10.1126/science.aag2302}.
\newblock \urlprefix\url{http://science.sciencemag.org/content/355/6325/602}

\bibitem{nomura_restricted_2017}
Y.~Nomura, A.S. Darmawan, Y.~Yamaji, M.~Imada, Physical Review B
  \textbf{96}(20), 205152 (2017).
\newblock \doi{10.1103/PhysRevB.96.205152}.
\newblock \urlprefix\url{https://link.aps.org/doi/10.1103/PhysRevB.96.205152}

\bibitem{Saito:2018b}
H.~Saito, Journal of the Physical Society of Japan \textbf{87}(7), 074002
  (2018).
\newblock \doi{10.7566/JPSJ.87.074002}.
\newblock \urlprefix\url{https://doi.org/10.7566/JPSJ.87.074002}

\bibitem{Choo:2018}
K.~Choo, G.~Carleo, N.~Regnault, T.~Neupert, Phys. Rev. Lett. \textbf{121},
  167204 (2018).
\newblock \doi{10.1103/PhysRevLett.121.167204}.
\newblock
  \urlprefix\url{https://link.aps.org/doi/10.1103/PhysRevLett.121.167204}

\bibitem{Nomura:2020}
Y.~Nomura, Journal of the Physical Society of Japan \textbf{89}(5), 054706
  (2020).
\newblock \doi{10.7566/JPSJ.89.054706}.
\newblock \urlprefix\url{https://journals.jps.jp/doi/10.7566/JPSJ.89.054706}

\bibitem{yoshioka:2019}
N.~Yoshioka, R.~Hamazaki, Physical Review B \textbf{99}(21), 214306 (2019).
\newblock \doi{10.1103/PhysRevB.99.214306}.
\newblock \urlprefix\url{https://link.aps.org/doi/10.1103/PhysRevB.99.214306}

\bibitem{nagy_variational_2019}
A.~Nagy, V.~Savona, Physical Review Letters \textbf{122}(25), 250501 (2019).
\newblock \doi{10.1103/PhysRevLett.122.250501}.
\newblock
  \urlprefix\url{https://link.aps.org/doi/10.1103/PhysRevLett.122.250501}

\bibitem{vicentini:2019}
F.~Vicentini, A.~Biella, N.~Regnault, C.~Ciuti, Physical Review Letters
  \textbf{122}(25), 250503 (2019).
\newblock \doi{10.1103/PhysRevLett.122.250503}.
\newblock
  \urlprefix\url{https://link.aps.org/doi/10.1103/PhysRevLett.122.250503}

\bibitem{hartmann_neural-network_2019}
M.J. Hartmann, G.~Carleo, Physical Review Letters \textbf{122}(25), 250502
  (2019).
\newblock \doi{10.1103/PhysRevLett.122.250502}.
\newblock
  \urlprefix\url{https://link.aps.org/doi/10.1103/PhysRevLett.122.250502}

\bibitem{ferrari_neural_2019}
F.~Ferrari, F.~Becca, J.~Carrasquilla, Physical Review B \textbf{100}(12),
  125131 (2019).
\newblock \doi{10.1103/PhysRevB.100.125131}.
\newblock \urlprefix\url{https://link.aps.org/doi/10.1103/PhysRevB.100.125131}.
\newblock Publisher: American Physical Society

\bibitem{Pfau:2019}
D.~{Pfau}, J.S. {Spencer}, A.e.G.d.G. {Matthews}, W.M.C. {Foulkes}, arXiv
  e-prints arXiv:1909.02487 (2019)

\bibitem{Hermann:2019}
J.~{Hermann}, Z.~{Sch{\"a}tzle}, F.~{No{\'e}}, arXiv e-prints arXiv:1909.08423
  (2019)

\bibitem{Choo:2019}
K.~Choo, A.~Mezzacapo, G.~Carleo, Nature Communications \textbf{11}(1), 2368
  (2020).
\newblock \doi{10.1038/s41467-020-15724-9}.
\newblock \urlprefix\url{https://www.nature.com/articles/s41467-020-15724-9}

\bibitem{Keeble:2019bkv}
J.W.T. Keeble, A.~Rios, Phys. Lett. B \textbf{809}, 135743 (2020).
\newblock \doi{10.1016/j.physletb.2020.135743}

\bibitem{Adams:2020aax}
C.~Adams, G.~Carleo, A.~Lovato, N.~Rocco, Phys. Rev. Lett. \textbf{127}(2),
  022502 (2021).
\newblock \doi{10.1103/PhysRevLett.127.022502}

\bibitem{Kievsky:2008jpg}
A.~Kievsky, S.~Rosati, M.~Viviani, L.~Marcucci, L.~Girlanda, J. Phys. G: Nucl.
  Part. Phys. \textbf{35}(6), 063101 (2008).
\newblock \doi{10.1088/0954-3899/35/6/063101}

\bibitem{Schiavilla:2021dun}
R.~Schiavilla, L.~Girlanda, A.~Gnech, A.~Kievsky, A.~Lovato, L.E. Marcucci,
  M.~Piarulli, M.~Viviani, Phys. Rev. C \textbf{103}(5), 054003 (2021).
\newblock \doi{10.1103/PhysRevC.103.054003}

\bibitem{Chen:1999tn}
J.W. Chen, G.~Rupak, M.J. Savage, Nucl. Phys. A \textbf{653}, 386 (1999).
\newblock \doi{10.1016/S0375-9474(99)00298-5}

\bibitem{Wiringa:1994wb}
R.B. Wiringa, V.G.J. Stoks, R.~Schiavilla, Phys. Rev. C \textbf{51}, 38 (1995).
\newblock \doi{10.1103/PhysRevC.51.38}

\bibitem{Yang:2019hkn}
C.J. Yang, Eur. Phys. J. A \textbf{56}(3), 96 (2020).
\newblock \doi{10.1140/epja/s10050-020-00104-0}

\bibitem{Bedaque:1998kg}
P.F. Bedaque, H.~Hammer, U.~van Kolck, Phys. Rev. Lett. \textbf{82}, 463
  (1999).
\newblock \doi{10.1103/PhysRevLett.82.463}

\bibitem{Metropolis:1953am}
N.~Metropolis, A.W. Rosenbluth, M.N. Rosenbluth, A.H. Teller, E.~Teller, J.
  Chem. Phys. \textbf{21}, 1087 (1953).
\newblock \doi{10.1063/1.1699114}

\bibitem{Massella:2018xdj}
P.~Massella, F.~Barranco, D.~Lonardoni, A.~Lovato, F.~Pederiva, E.~Vigezzi, J.
  Phys. G \textbf{47}, 035105 (2020).
\newblock \doi{10.1088/1361-6471/ab588c}

\bibitem{Zaheer:2017}
M.~{Zaheer}, S.~{Kottur}, S.~{Ravanbakhsh}, B.~{Poczos}, R.~{Salakhutdinov},
  A.~{Smola}, arXiv e-prints arXiv:1703.06114 (2017)

\bibitem{Wagstaff:2019}
E.~{Wagstaff}, F.B. {Fuchs}, M.~{Engelcke}, I.~{Posner}, M.~{Osborne}, arXiv
  e-prints arXiv:1901.09006 (2019)

\bibitem{Dugas:2000}
C.~Dugas, Y.~Bengio, F.~Bélisle, C.~Nadeau, R.~Garcia, in \emph{Advances in
  {Neural} {Information} {Processing} {Systems} 13}, ed. by T.K. Leen, T.G.
  Dietterich, V.~Tresp (MIT Press, 2001), pp. 472--478.
\newblock
  \urlprefix\url{http://papers.nips.cc/paper/1920-incorporating-second-order-functional-knowledge-for-better-option-pricing.pdf}

\bibitem{Sorella:2005}
S.~Sorella, Phys. Rev. B \textbf{71}, 241103 (2005).
\newblock \doi{10.1103/PhysRevB.71.241103}.
\newblock \urlprefix\url{http://link.aps.org/doi/10.1103/PhysRevB.71.241103}

\bibitem{Marcucci:2019fphy}
L.E. Marcucci, J.~Dohet-Eraly, L.~Girlanda, A.~Gnech, A.~Kievsky, M.~Viviani,
  Front. Phys. \textbf{8}, 69 (2020).
\newblock \doi{10.3389/fphy.2020.00069}.
\newblock
  \urlprefix\url{https://www.frontiersin.org/article/10.3389/fphy.2020.00069}

\bibitem{Gnech:2020prc}
A.~Gnech, M.~Viviani, L.E. Marcucci, Phys. Rev. C \textbf{102}, 014001 (2020).
\newblock \doi{10.1103/PhysRevC.102.014001}.
\newblock \urlprefix\url{https://link.aps.org/doi/10.1103/PhysRevC.102.014001}

\bibitem{Gnech:2021prc}
A.~Gnech, L.E. Marcucci, R.~Schiavilla, M.~Viviani,   (2021).
\newblock ArXiv:2016.07439. To be published on Phys. Rev. C

\bibitem{ParticleDataGroup:2012pjm}
J.~Beringer, et~al., Phys. Rev. D \textbf{86}, 010001 (2012).
\newblock \doi{10.1103/PhysRevD.86.010001}

\bibitem{Friar:1997js}
J.L. Friar, J.~Martorell, D.W.L. Sprung, Phys. Rev. A \textbf{56}, 4579 (1997).
\newblock \doi{10.1103/PhysRevA.56.4579}

\bibitem{NNDC2021}
NNDC (2021), Nudat2, https://www.nndc.bnl.gov/nudat2/

\bibitem{CODATA2021}
E.~Tiesinga, P.J. Mohr, D.B. Newell, B.N. Taylor, CODATA2018  (2020).
\newblock Available at http://physics.nist.gov/constants

\bibitem{Amroun:1994npa}
A.~Amroun, V.~Breton, J.M. Cavedon, B.~Frois, D.~Goutte, F.~Juster, P.~Leconte,
  J.~Martino, Y.~Mizuno, X.H. Phan, S.~Platchkov, I.~Sick, S.~Williamson,
  Nuclear Physics A \textbf{579}(3), 596 (1994).
\newblock \doi{https://doi.org/10.1016/0375-9474(94)90925-3}.
\newblock
  \urlprefix\url{https://www.sciencedirect.com/science/article/pii/0375947494909253}

\bibitem{Morton:2006pra}
D.C. Morton, Q.~Wu, G.W.F. Drake, Phys. Rev. A \textbf{73}, 034502 (2006).
\newblock \doi{10.1103/PhysRevA.73.034502}.
\newblock \urlprefix\url{https://link.aps.org/doi/10.1103/PhysRevA.73.034502}

\bibitem{Krauth:2021nat}
J.~Krauth, K.~Schuhmann, M.~Ahmed, {\it et al.}, Nature \textbf{598}, 527
  (2021).
\newblock \doi{10.1038/s41586-021-03183-1}

\bibitem{Wang:2004prl}
L.B. Wang, P.~Mueller, K.~Bailey, G.W.F. Drake, J.P. Greene, D.~Henderson, R.J.
  Holt, R.V.F. Janssens, C.L. Jiang, Z.T. Lu, T.P. O'Connor, R.C. Pardo, K.E.
  Rehm, J.P. Schiffer, X.D. Tang, Phys. Rev. Lett. \textbf{93}, 142501 (2004).
\newblock \doi{10.1103/PhysRevLett.93.142501}.
\newblock
  \urlprefix\url{https://link.aps.org/doi/10.1103/PhysRevLett.93.142501}

\bibitem{Puchalski:2013prl}
M.~Puchalski, K.~Pachucki, Phys. Rev. Lett. \textbf{111}, 243001 (2013).
\newblock \doi{10.1103/PhysRevLett.111.243001}.
\newblock
  \urlprefix\url{https://link.aps.org/doi/10.1103/PhysRevLett.111.243001}

\bibitem{Wiringa:2021p}
R.~Wiringa.
\newblock Private communication

\bibitem{Contessi:2017rww}
L.~Contessi, A.~Lovato, F.~Pederiva, A.~Roggero, J.~Kirscher, U.~van Kolck,
  Phys. Lett. B \textbf{772}, 839 (2017).
\newblock \doi{10.1016/j.physletb.2017.07.048}

\bibitem{Kamada:2001tv}
H.~Kamada, et~al., Phys. Rev. C \textbf{64}, 044001 (2001).
\newblock \doi{10.1103/PhysRevC.64.044001}

\bibitem{Maris:2013rgq}
P.~Maris, J.P. Vary, S.~Gandolfi, J.~Carlson, S.C. Pieper, Phys. Rev. C
  \textbf{87}(5), 054318 (2013).
\newblock \doi{10.1103/PhysRevC.87.054318}

\bibitem{Furnstahl:2014xsa}
R.J. Furnstahl, D.R. Phillips, S.~Wesolowski, J. Phys. G \textbf{42}(3), 034028
  (2015).
\newblock \doi{10.1088/0954-3899/42/3/034028}

\bibitem{Melendez:2019izc}
J.A. Melendez, R.J. Furnstahl, D.R. Phillips, M.T. Pratola, S.~Wesolowski,
  Phys. Rev. C \textbf{100}(4), 044001 (2019).
\newblock \doi{10.1103/PhysRevC.100.044001}

\end{thebibliography}

\end{document}